\documentclass[preprint2]{aastex}
\usepackage{spr-astr-addons}
\usepackage{url}
\bibpunct{(}{)}{,}{n}{,}{,}

\RequirePackage{color}

\begin{document}

 \title{From horseshoe to quasi-satellite and back again:
        the curious dynamics of Earth co-orbital asteroid 
        2015~SO$_{2}$ 
        }

 \shorttitle{Dynamical evolution of 2015~SO$_{2}$}
 \shortauthors{de la Fuente Marcos and de la Fuente Marcos}

 \author{C.~de~la~Fuente Marcos} 
  \and 
 \author{R.~de~la~Fuente Marcos} 
 \affil{Apartado de Correos 3413, E-28080 Madrid, Spain} 
 \email{carlosdlfmarcos@gmail.com}

 \begin{abstract}
    Earth co-orbitals of the horseshoe type are interesting
    objects to study for practical reasons. They are 
    relatively easy to access from our planet and that makes 
    them attractive targets for sample return missions. Here, 
    we show that near-Earth asteroid (NEA) 2015~SO$_{2}$ is 
    a transient co-orbital to the Earth that experiences a 
    rather peculiar orbital evolution characterised by 
    recurrent, alternating horseshoe and quasi-satellite 
    episodes. It is currently following a horseshoe 
    trajectory, the ninth asteroid known to do so. Besides 
    moving inside the 1:1 mean motion resonance with the 
    Earth, it is subjected to a Kozai resonance with the 
    value of the argument of perihelion librating around 
    270\degr. Contrary to other NEAs, asteroid 2015~SO$_{2}$ 
    may have remained in the vicinity of Earth's co-orbital 
    region for a few hundreds of thousands of years. 
 \end{abstract}


 \section{Introduction}
    In the Solar System, co-orbital asteroids share the orbit of a planet, going around the Sun in almost exactly one orbital period of 
    their host planet, i.e., move inside the 1:1 mean motion resonance with a planet. These minor bodies are not truly captured by the 
    gravity of the host planet as natural satellites are, but from the planet's point of view they loop around, sometimes in remarkably 
    stable orbits. 

    Co-orbital motion is characterised by a suitable angular variable, the resonant or critical angle, that in this case is the difference 
    between the mean longitudes of the asteroid and its host planet or relative mean longitude. The mean longitude of an object ---planet or 
    minor body--- is given by $\lambda=M+\Omega+\omega$, where $M$ is the mean anomaly, $\Omega$ is the longitude of the ascending node, and 
    $\omega$ is the argument of perihelion (see, e.g., Murray \& Dermott 1999). For these objects, the value of the critical angle librates 
    or oscillates; for a regular passing body, the relative mean longitude circulates or takes any value in the range (0, $2\pi$). Even if
    some of them can experience relatively close flybys, co-orbital asteroids tend to be small, transient visitors and they do not pose a 
    particularly high hazard for their host planets. 

    Dynamically speaking, the co-orbital state is surprisingly rich and it includes three elementary configurations ---quasi-satellite or 
    retrograde satellite, Trojan or tadpole, and horseshoe--- and a multiplicity of hybrid arrangements or compound orbits (for instance, 
    combination of quasi-satellite and tadpole orbits, Morais \& Morbidelli 2002). Recurrent transitions between the main resonant states 
    are possible (Namouni et al. 1999; Namouni \& Murray 2000). The quasi-satellite dynamical state is observed when the value of the 
    relative mean longitude, $\lambda_{\rm r}$, oscillates around 0\degr (see, e.g., Mikkola et al. 2006). If the libration is around 
    $\pm60\degr$ we call the object a Trojan and it follows a tadpole orbit (see, e.g., Murray \& Dermott 1999); such an object is 
    classified as an $L_4$ Trojan when the value of $\lambda_{\rm r}$ librates around +60\degr, or as an $L_5$ Trojan if the value 
    oscillates around $-60\degr$ (or 300\degr). The Trojan configuration is the most prevalent co-orbital state among long-term, stable 
    co-orbitals. However, the most usual co-orbital dynamical state among transient co-orbitals is the one characterised by a libration 
    amplitude larger than 180\degr around a value of $\lambda_{\rm r}=180\degr$ and often enclosing $\pm60\degr$; such objects follow 
    horseshoe orbits (see, e.g., Murray \& Dermott 1999). At regular intervals and before moving away, a horseshoe librator follows a 
    corkscrew-like trajectory in the vicinity of its host planet for a certain period of time, a decade or more in the case of the Earth; 
    these episodes are recurrent.

    Hollabaugh and Everhart (1973) predicted the existence of asteroids moving in long-lasting horseshoe orbits associated with the Earth. 
    None was found until 3753 Cruithne (1986 TO) was identified by Wiegert et al. (1997) as an asteroidal companion to our planet following
    a horseshoe orbit. Nearly two decades later, the number of known asteroids which are Earth co-orbitals remains relatively small. Our 
    planet hosts one Trojan ---2010~TK$_{7}$ (Connors et al. 2011)---, four quasi-satellites ---164207 (2004~GU$_{9}$) (Connors et al. 2004; 
    Mikkola et al. 2006; Wajer 2010), 277810 (2006~FV$_{35}$) (Wiegert et al. 2008; Wajer 2010), 2013~LX$_{28}$ (Connors 2014), and 
    2014~OL$_{339}$ (de la Fuente Marcos \& de la Fuente Marcos 2014)---, and eight horseshoe librators ---3753 (Wiegert et al. 1997, 1998), 
    85770 (1998~UP$_{1}$),\footnote{\url{http://www.astro.uwo.ca/~wiegert/eca/}} 54509 YORP (2000 PH$_{5}$) (Wiegert et al. 2002; Margot \& 
    Nicholson 2003), 2001~GO$_{2}$ (Wiegert et al. 2002; Margot \& Nicholson 2003; Brasser et al. 2004), 2002~AA$_{29}$ (Connors et al. 
    2002; Brasser et al. 2004), 2003~YN$_{107}$ (Brasser et al. 2004; Connors et al. 2004), 2010~SO$_{16}$ (Christou \& Asher 2011), and 
    2013~BS$_{45}$ (de la Fuente Marcos \& de la Fuente Marcos 2013). Most of these objects are not large enough to cause widespread damage 
    if they strike our planet, but there are some outliers. While objects such as 2003~YN$_{107}$, 2002~AA$_{29}$ or 2001~GO$_{2}$ are 
    probably not large enough to reach the ground ---in the rare event of a collision--- with a significant fraction of their original 
    kinetic energy remaining, others, e.g., 3753 ($H=15.7$~mag) or 85770 ($H=20.5$~mag) are significantly larger than either the Tunguska 
    (see, e.g., Yeomans 2006) or Chelyabinsk (Brown et al. 2013) impactors and may cause considerable destruction over populated and/or 
    coastal areas. Based on their current orbital solutions, the 13 objects cited are present-day ---but transient--- Earth co-orbital 
    asteroids; the values of their respective $\lambda_{\rm r}$ librate or oscillate as described above although some of them are in the 
    process of transitioning between co-orbital states or follow compound orbits. These co-orbitals are often considered rare curiosities, 
    but their dynamical behaviour and affordable accessibility from the Earth (e.g. Stacey \& Connors 2009) make them very good candidates 
    for future interplanetary space activities such as in situ study, sample return missions, or even commercial mining (e.g. Lewis 1996; 
    Elvis 2012, 2014; Garc\'{\i}a Y\'arnoz et al. 2013; Harris \& Drube 2014). 

    Here, we show that the recently discovered asteroid 2015~SO$_{2}$ is performing the corkscrew motion characteristic of Earth co-orbitals
    of the horseshoe type. This object is the 14th known Earth co-orbital and the 9th horseshoe librator. This paper is organised as 
    follows. In Sect. 2, we present the available data on 2015~SO$_{2}$ and the methodology followed in this study. The dynamical evolution 
    of this minor body is studied in Sect. 3. The details of the mechanism triggering the transitions between the horseshoe and the 
    quasi-satellite dynamical states are explored in Sect. 4. In Sect. 5, we consider the impact of errors on our results. A discussion is 
    presented in Sect. 6 and in Sect. 7 we summarise our conclusions.

 \section{Asteroid 2015~SO$_{2}$: data and methodology}
    Asteroid 2015~SO$_{2}$ was discovered on 2015 September 21 by B. Miku\v{z} and S. Maticic observing with the 0.6-m f/3.3 Cichocki
    telescope at the \v{C}rni Vrh Observatory in Slovenia (Miku\v{z} et al. 2015). It had a $R$ magnitude of 19.4 when first observed. It is 
    a small object with $H$ = 23.9 mag, which translates into a diameter in the range 50--111 m for an assumed albedo of 0.20--0.04. Its 
    orbit is moderately well determined with 84 observations acquired during 9 d (see Table \ref{elements}) and it is typical of a minor 
    body that moves co-orbitally with the Earth. The source of the Heliocentric Keplerian osculating orbital elements and uncertainties in 
    Table \ref{elements} is the JPL Small-Body Database\footnote{\url{http://ssd.jpl.nasa.gov/sbdb.cgi}} and they are referred to the epoch
    JD2457200.5, i.e. a time prior to its discovery and subsequent close encounter with our planet. At that time, the object was an Apollo
    asteroid moving in an orbit with a value of the semi-major axis $a$ = 1.00079 AU, very close to that of our planet (0.99957 AU), 
    relatively low eccentricity, $e$ = 0.11, and moderate inclination, $i$ = 9\fdg2. Since the close encounter with the Earth, 2015~SO$_{2}$ 
    has become an Aten asteroid.
%
%
     \begin{table}
       \centering
        \fontsize{8}{11pt}\selectfont
        \tabcolsep 0.10truecm
        \caption{Heliocentric ecliptic Keplerian orbital elements of asteroid 2015~SO$_{2}$. Values include the 1$\sigma$ uncertainty (Epoch 
                 = JD2457200.5, 2015-June-27.0; J2000.0 ecliptic and equinox. Source: JPL Small-Body Database.)
                }
        \begin{tabular}{lll}
         \hline
          Parameter                                         &   &                 Value \\
         \hline
          Semi-major axis, $a$ (AU)                         & = &   1.00079$\pm$0.00004 \\
          Eccentricity, $e$                                 & = &   0.10758$\pm$0.00009 \\
          Inclination, $i$ (\degr)                          & = &   9.198$\pm$0.010     \\
          Longitude of the ascending node, $\Omega$ (\degr) & = & 183.029$\pm$0.002     \\
          Argument of perihelion, $\omega$ (\degr)          & = & 289.28$\pm$0.02       \\
          Mean anomaly, $M$ (\degr)                         & = & 174.47$\pm$0.03       \\
          Perihelion, $q$ (AU)                              & = &   0.89312$\pm$0.00005 \\
          Aphelion, $Q$ (AU)                                & = &   1.10845$\pm$0.00005 \\
          Absolute magnitude, $H$ (mag)                     & = &  23.9                 \\
         \hline
        \end{tabular}
        \label{elements}
     \end{table}
%
%

    In the following, we study the short-term dynamical evolution of this recently discovered near-Earth asteroid (NEA). Our analysis is 
    based on results of direct $N$-body calculations that use the most updated ephemerides and include perturbations from the eight major 
    planets, the Moon, the barycentre of the Pluto-Charon system, and the three largest asteroids. The extensive numerical integrations
    presented here have been performed using the Hermite scheme described by Makino (1991) and implemented by Aarseth (2003). The standard 
    version of this direct $N$-body code is publicly available from the IoA web site.\footnote{\url{http://www.ast.cam.ac.uk/~sverre/web/pages/nbody.htm}}
    Non-gravitational forces, relativistic and oblateness terms are not included in the calculations; for further details, see de la 
    Fuente Marcos \& de la Fuente Marcos (2012). Initial conditions (positions and velocities in the barycentre of the Solar System) have 
    been obtained from the Jet Propulsion Laboratory (JPL) HORIZONS system (Giorgini et al. 1996; Standish 1998) and they are referred to 
    as the JD 2457200.5 epoch (2015-June-27.0), which is the $t$ = 0 instant in our figures.

    Although the current orbit of 2015~SO$_{2}$ is relatively poor, if enough test orbits are studied, the best estimates of the past and 
    future dynamical evolution of this object can be determined. This assumption is based on the widely accepted idea that statistical 
    results of an ensemble of collisional $N$-body simulations are accurate, even though individual simulations are not (see, e.g., Boekholt 
    \& Portegies Zwart 2015). For values of the standard deviation of the orbital elements as moderate as those in Table \ref{elements}, if 
    consistent behaviour is systematically found within reasonable limits, the dynamical nature of the object can be firmly established. In 
    addition to the calculations completed using the nominal orbital elements in Table \ref{elements}, we have performed 50 control 
    simulations with sets of orbital elements obtained from the nominal ones within the accepted error limits (up to 9$\sigma$) that reflect 
    the observational uncertainty in astrometry (see Sect. 5 for details). In the figures when an orbit is labelled `$\pm{n}\sigma$', where 
    $n$ is an integer, it has been obtained by adding (+) or subtracting ($-$) $n$-times the uncertainty from the orbital parameters (the 
    six elements) in Table \ref{elements}. All the control orbits exhibit consistent behaviour within a few hundred years of $t = 0$, which 
    is the Lyapunov time ---time-scale for exponential divergence of initially close orbits--- for this object.

 \section{Dynamical evolution}
\begin{figure}
        \centering
         \includegraphics[width=\linewidth]{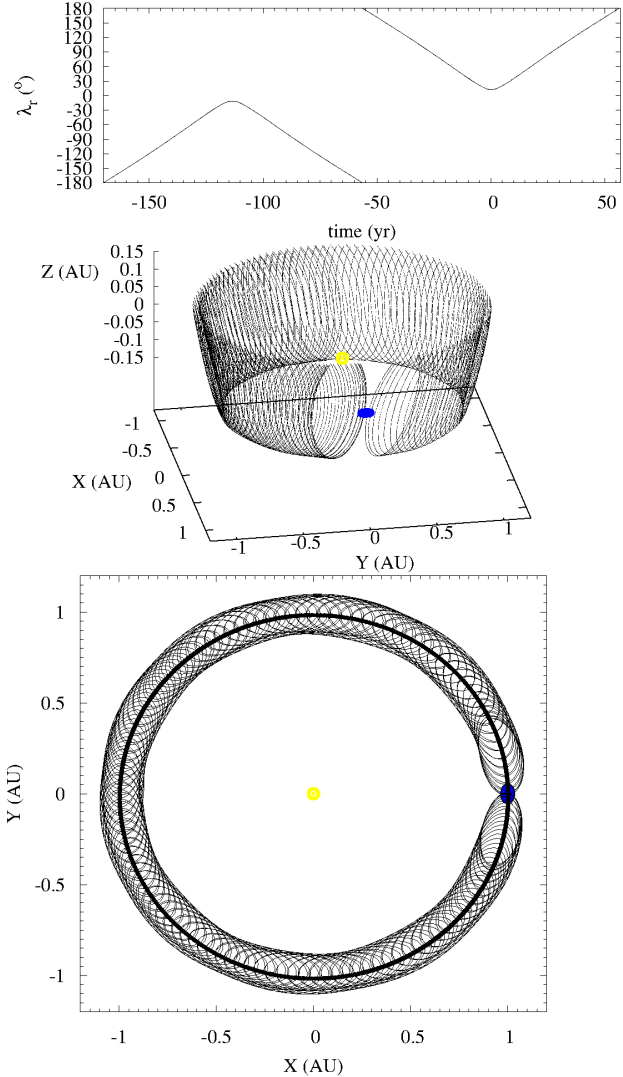}
         \caption{Orbital evolution of 2015~SO$_{2}$ in the time interval (-170, 57) yr as described by the nominal orbit in Table 
                  \ref{elements}. Evolution of the relative mean longitude, $\lambda_{\rm r}$, during one period of the horseshoe orbit (top 
                  panel). The motion of 2015~SO$_{2}$ in a coordinate system rotating with the Earth, in space (middle panel) and displayed 
                  projected onto the ecliptic plane (bottom panel). The orbit and position of our planet are also indicated. All the 
                  investigated control orbits ($\pm9\sigma$) exhibit the same behaviour within this timeframe.
                 }
         \label{orbit}
      \end{figure}

    In order to study the dynamics of 2015~SO$_{2}$ we have performed $N$-body calculations in both directions of time for 50 kyr using the 
    physical model described above. Figure \ref{orbit}, top panel, shows the evolution of the relative mean longitude, $\lambda_{\rm r}$,
    during one period of the horseshoe orbit. The value of $\lambda_{\rm r}$ behaves as expected of a classical horseshoe librator (see, 
    e.g., Murray \& Dermott 1999). The motion of 2015~SO$_{2}$ over the time range (-170, 57) yr as seen in a coordinate system rotating 
    with the Earth in space (middle panel) and projected onto the ecliptic plane (bottom panel) is plotted in Fig. \ref{orbit} (nominal 
    orbit in Table \ref{elements}). Asteroid 2015~SO$_{2}$ is an Earth co-orbital; it is performing the corkscrew motion characteristic of 
    Earth co-orbitals of the horseshoe type (see Fig. \ref{orbit}, middle panel). All the investigated control orbits (within $\pm9\sigma$) 
    exhibit the same behaviour within the timeframe mentioned above. Based solely on the number of simulations performed, we estimate the 
    probability of 2015~SO$_{2}$ being a present-day horseshoe librator to the Earth at $>99.9$\%. The motion in Fig. \ref{orbit} is the 
    result of the superposition of a $\sim1$~yr epicyclic motion describing a kidney-like path and a slow oscillation in mean longitude with 
    a half period close to 113 yr (the time to reach the other end of the horseshoe). The shape of the epicycle-like path is affected by the 
    value of the eccentricity of the orbit. Minor bodies subjected to this type of dynamics may have had their origin in the Earth-Moon 
    system (see, e.g., Margot \& Nicholson 2003) although other sources in the main asteroid belt are possible. As pointed out by Connors et 
    al. (2004) the effects derived from orbital chaos severely limit our ability to determine the source of these NEAs.
%
%
%
%

\begin{figure}
       \centering
        \includegraphics[width=\linewidth]{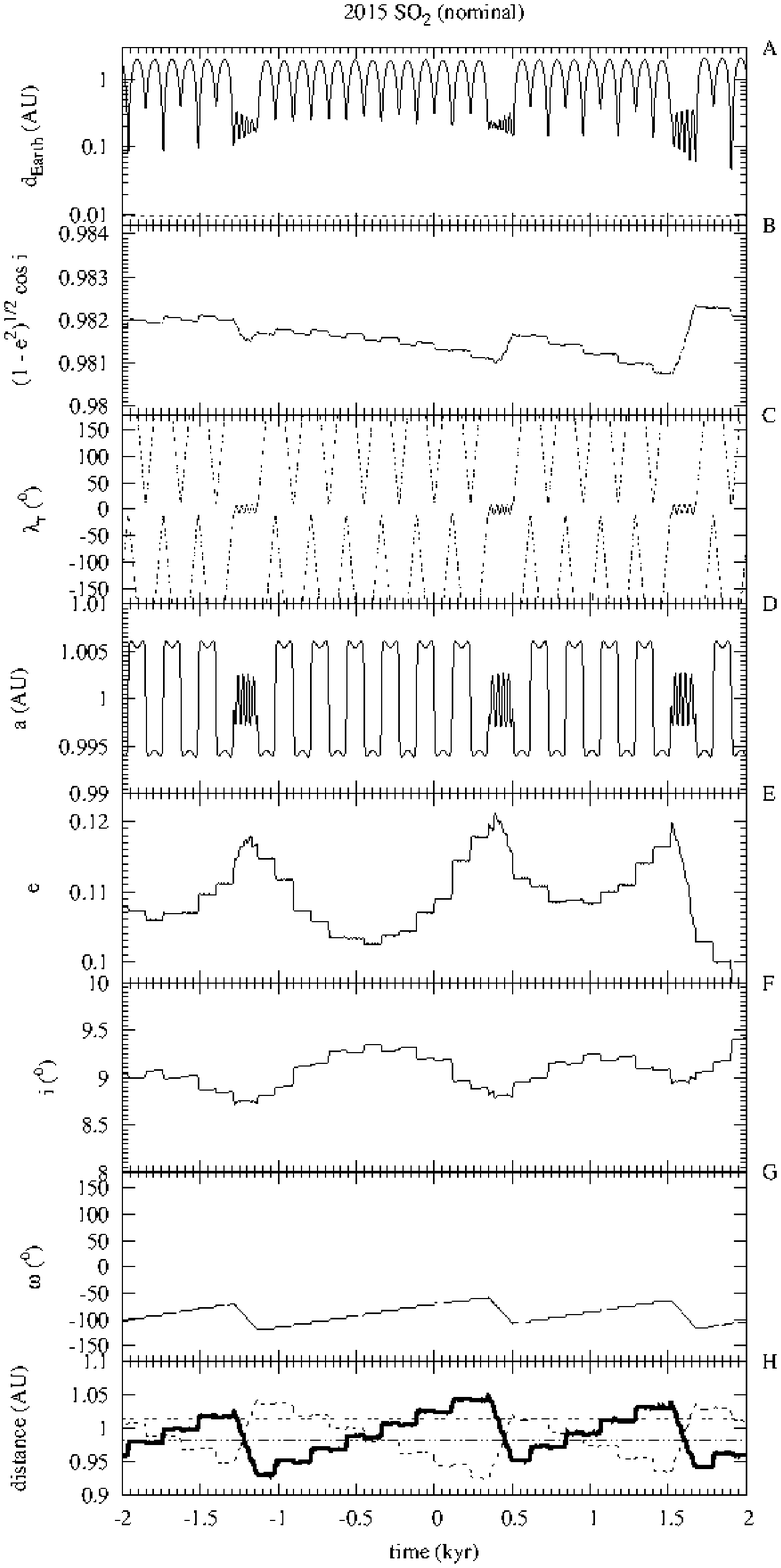}
        \caption{Time evolution of various parameters for the nominal orbit during the time interval (-2000, 2000) yr. The distance from the 
                 Earth (panel A) with the value of the radius of the Hill sphere of the Earth, 0.0098 AU (dashed line). The parameter 
                 $\sqrt{1 - e^2} \cos i$ (panel B). The resonant angle, $\lambda_{\rm r}$ (panel C). The orbital elements $a$ (panel D), $e$ 
                 (panel E), $i$ (panel F), and $\omega$ (panel G). The distances to the descending (thick line) and ascending nodes (dotted 
                 line) are plotted in panel H; Earth's aphelion and perihelion distances are also shown.
                }
        \label{short}
     \end{figure}

    The short-term dynamical evolution of the nominal orbit in Table \ref{elements} is presented in Fig. \ref{short}. Figure \ref{short}, 
    panel D, shows that 2015~SO$_{2}$, like other horseshoe librators, alternates between the Aten and Apollo dynamical classes. Prior to 
    its close encounter with the Earth on 2015 September 30, Earth's gravity was decreasing the asteroids's orbital energy in such a way 
    that its semi-major axis was changing from $\sim1.006$~AU (Apollo-type orbit) to a value of $\sim0.994$~AU (Aten-type orbit). As an 
    Aten, it will reach the other end of the horseshoe path in about 113 years from now, going from the leading side of the Earth to the
    trailing side. At that time, Earth's gravity will increase the asteroid's orbital energy so its current value of the semi-major axis of 
    $\sim0.994$~AU will become $\sim1.006$~AU. Figure \ref{short}, panel A, shows that 2015~SO$_{2}$ has only experienced relatively distant 
    close encounters with our planet during the time interval displayed (4000~yr), well beyond the value of the radius of the Hill sphere of 
    the Earth, 0.0098 AU. The figure also shows that although 2015~SO$_{2}$ currently follows a horseshoe orbit with respect to the Earth, 
    it will become a quasi-satellite of our planet (the value of $\lambda_{\rm r}$ librating around 0\degr) in about 350 yr (see panel C). 
    This dynamical state will persist for about 150 yr. 

    Figure \ref{short}, panel G, shows that the argument of perihelion of 2015~SO$_{2}$ oscillates around $-90\degr$ or 270\degr. Therefore,
    this object reaches perihelion when it is farthest from the ecliptic plane, i.e., close encounters with inner planets are not possible
    at perihelion or aphelion. When an object exhibits this behaviour, it is said to be affected by the Kozai mechanism or trapped in a 
    Kozai resonance (Kozai 1962). Because of the Kozai resonance, both eccentricity and inclination oscillate with the same frequency but 
    out of phase (see Fig. \ref{short}, panels E and F); when the value of the eccentricity reaches its maximum the value of the inclination 
    is the lowest and vice versa ($\sqrt{1 - e^2} \cos i \sim$ constant, see Fig. \ref{short}, panel B). Although certainly present as 
    verified by the behaviour of the invariant, the Kozai resonance is not very strong.
%
%
%
%

\begin{figure*}
       \centering
        \includegraphics[width=\linewidth]{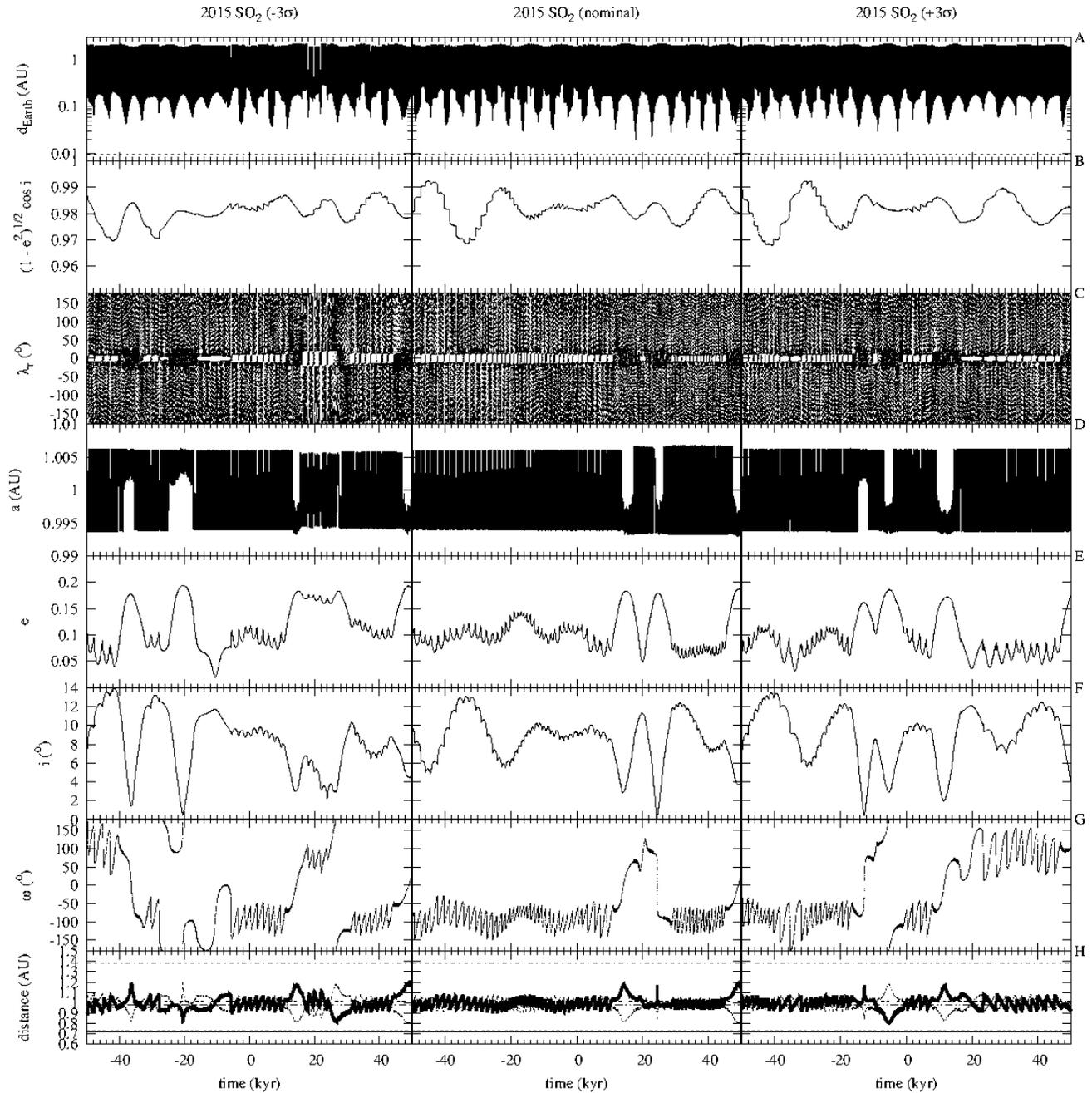}
        \caption{Same as Fig. \ref{short} but for the nominal orbit (same data as in Fig. \ref{short} but over a longer time span) and two 
                 representative examples of orbits that are the most different from the nominal one among those integrated up to 
                 $\pm3\sigma$ deviations (see text for details). Venus' and Mars' aphelion and perihelion distances are also shown in panel 
                 H.
                }
        \label{first}
     \end{figure*}

    Figure \ref{first}, central panels, shows the results of the entire integration. We confirm that 2015~SO$_{2}$ does not experience 
    particularly close encounters with the Earth; a large number of transitions from horseshoe to quasi-satellite and back again are 
    observed. The object exhibits Kozai-like behaviour throughout the entire time interval with its nodes confined between Earth's aphelion 
    and perihelion most of the time. The distance between the Sun and the nodes for a prograde orbit is given by
    \begin{equation}
       r=a(1-e^2)/(1\pm{e}\cos\omega)\,, \label{nodeseq}
    \end{equation}
    where the "+" sign is for the ascending node (where the orbit crosses the Ecliptic from South to North) and the "$-$" sign is for the
    descending node. Longer integrations not shown here indicate that it may have reached Earth's co-orbital region 500--200 kyr ago and it 
    could leave in another 100--200 kyr to follow orbits interior to that of the Earth. Based on the integrations performed, the Earth's 
    co-orbital region extends from $\sim0.994$~AU to $\sim1.006$~AU.
%
%
%
%

 \section{From horseshoe to quasi-satellite and back again: the mechanism}
    Figure \ref{short}, panel C, shows that in general 2015~SO$_{2}$ follows a horseshoe orbit with respect to our planet but has periods of
    quasi-satellite behaviour; in 4000 years of simulated time we observe three instances of the quasi-satellite phase with an average 
    duration of 150 yr per episode. Asteroid 2015~SO$_{2}$ is not the first transient Earth co-orbital of the horseshoe type that suffers 
    repeated transitions to the quasi-satellite dynamical state, Brasser et al. (2004) have documented this behaviour for 2001~GO$_{2}$, 
    2002~AA$_{29}$ and 2003~YN$_{107}$. These transitions had been predicted and explained by Namouni (1999) and Christou (2000), and depend 
    on the influence of other planets, in particular the giant planets and Venus, owing to the overlapping of multiple secular resonances. 
    This is not surprising as Ito \& Tanikawa (1999) pointed out that the terrestrial planets share the effect of the secular perturbation 
    from Jupiter. In a follow-up work, Tanikawa \& Ito (2007) further extended this analysis concluding that, regarding the secular 
    perturbation from Jupiter, the inner planets are a planetary group or collection of loosely connected mutually dynamically dependent 
    planets. The effects of the overlapping secular resonances persist even outside the region where $\lambda_{\rm r}$ librates and the 
    dynamics of some passing bodies ---which are Kozai librators because their argument of perihelion oscillates--- is still controlled by 
    them (de la Fuente Marcos \& de la Fuente Marcos 2015a). Brasser et al. (2004) noticed that, in the cases studied by them, the 
    transition to the quasi-satellite phase was possible both at the leading and at the trailing sides of the Earth, although most 
    transitions took place at the trailing side of our planet.

    Figure \ref{short}, panel C, shows that quasi-satellite episodes start when the nodes are farthest from each other (see panel H) and 
    also farthest from the path of the Earth. In the Solar System and for a minor body moving in an inclined orbit, like 2015~SO$_{2}$, 
    close encounters with major planets are only possible in the vicinity of the nodes. Therefore, the start of the quasi-satellite phase of 
    the co-orbital motion coincides with the time when the average gravitational effects of the Earth on the object are the weakest possible. 
    The transition back to the horseshoe phase is triggered when both nodes are close to Earth's perihelion; then, encounters with the 
    Earth-Moon system are possible at both nodes and the sustained action of these encounters increases the asteroid's orbital energy 
    eventually triggering the transition and returning the minor body to a horseshoe trajectory. In our simulations, the transition from 
    horseshoe librator to quasi-satellite always takes place when the orbital energy of the asteroid increases after having followed a 
    trajectory of the Aten type for nearly 100 yr. In other words, all the observed transitions from horseshoe to quasi-satellite took place
    at the trailing side of the Earth, never at the leading side. During the quasi-satellite phase, the orbital eccentricity is the highest 
    and the inclination, the lowest; the value of the argument of perihelion decreases as predicted by Namouni (1999). Also consistent with
    Namouni (1999), during the transitions from horseshoe to quasi-satellite, the argument of perihelion is stationary and a maximum; the
    value of the eccentricity must be higher than a critical one to start the transition. Transitions are only observed when the descending
    node is beyond Earth's aphelion and the ascending node is interior to Earth's perihelion which is likely the result of overlapping 
    secular resonances. 

 \section{Impact of errors on the short-term dynamical evolution}
\begin{figure*}
       \centering
        \includegraphics[width=\linewidth]{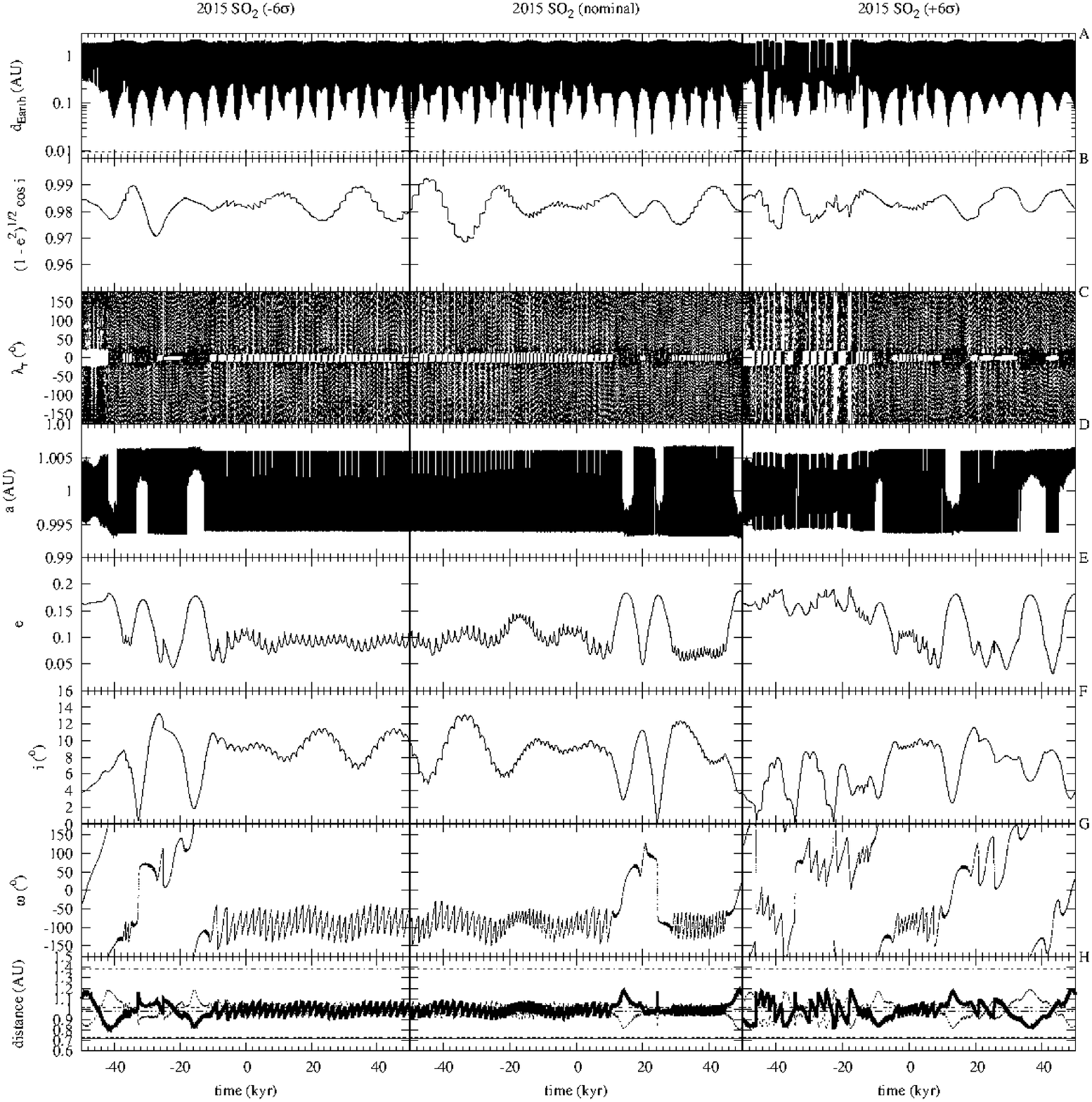}
        \caption{Same as Fig. \ref{first} but for the nominal orbit (same data as in Fig. \ref{first}) and two representative examples of 
                 orbits that are the most different from the nominal one among those integrated up to $\pm6\sigma$ deviations (see text for 
                 details).
                }
        \label{second}
     \end{figure*}

\begin{figure*}
       \centering
        \includegraphics[width=\linewidth]{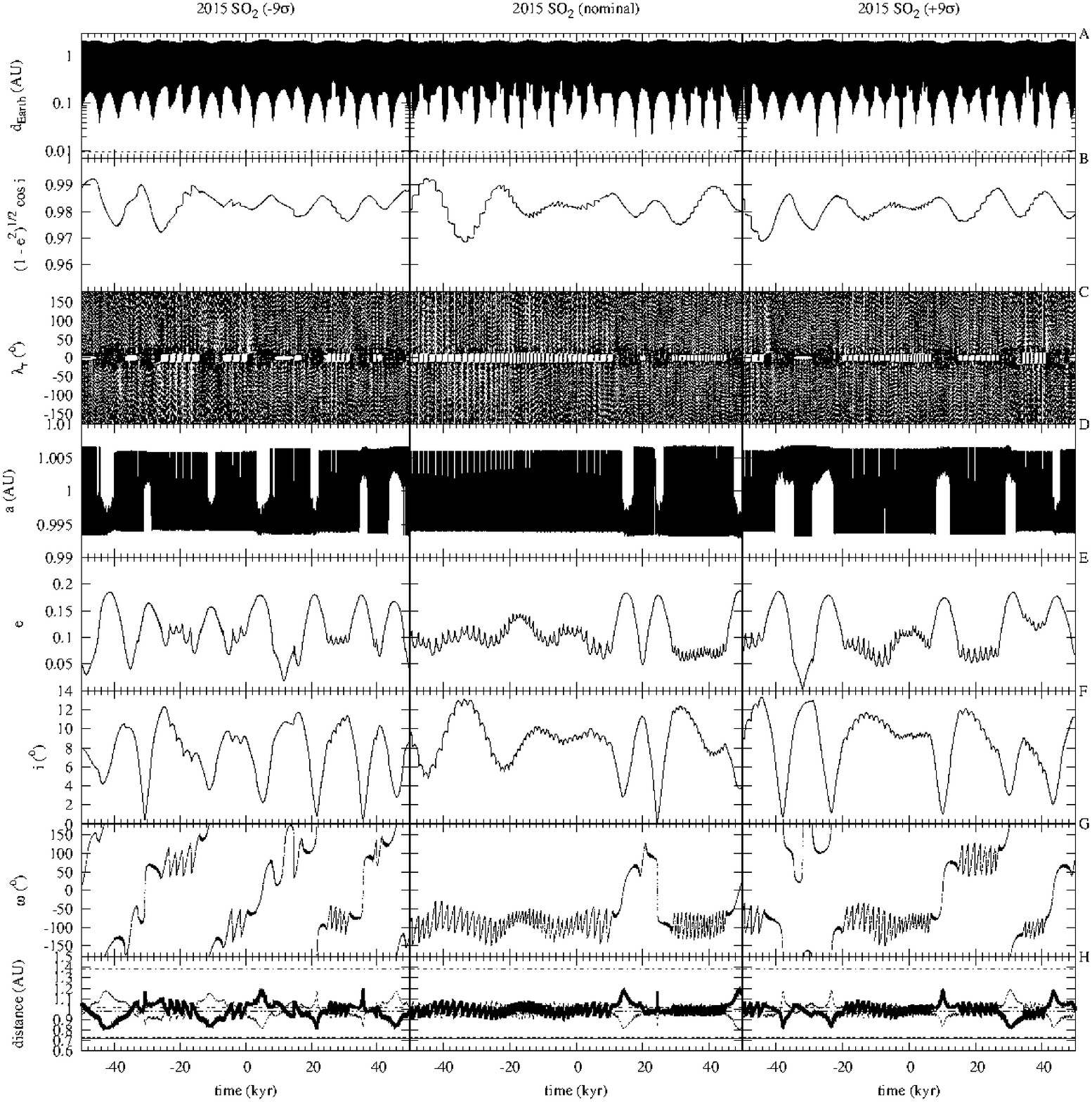}
        \caption{Same as Fig. \ref{first} but for the nominal orbit (same data as in Fig. \ref{first}) and two representative examples of 
                 orbits that are the most different from the nominal one among those integrated up to $\pm9\sigma$ deviations (see text for 
                 details).
                }
        \label{third}
     \end{figure*}

    It may be argued that the orbit of 2015~SO$_{2}$ currently available is based on a short arc and, therefore, this makes our conclusions 
    rather weak or even entirely questionable. However, not all orbital solutions with short data-arc spans are equally poor. The object 
    discussed here follows a relatively stable orbit that is only directly perturbed by the Earth--Moon system. In addition to the 
    integrations performed making use of the nominal orbital parameters in Table \ref{elements}, we have computed 50 control simulations 
    with sets of orbital elements obtained from the nominal ones within the quoted uncertainties and assuming Gaussian distributions for 
    them up to $\pm9\sigma$. Representative results of these calculations are displayed in Figs. \ref{first}, \ref{second} and \ref{third}. 
    The orbital evolution of 2015~SO$_{2}$ within a few thousand years of $t=0$ is not too different from that in Fig. \ref{short} even if
    deviations in the values of the orbital elements as large as $\pm3\sigma$ or even $\pm6\sigma$ are considered. For the extreme case of
    $\pm9\sigma$, the overall dynamical behaviour is still consistent: the object remains within Earth's co-orbital zone for the entire
    integration. These results clearly indicate that our conclusions are robust and reliable.
%
%
%
%
%
%
%
%

    \subsection{Impact of errors: classical treatment}
       A more detailed analysis of the effects of errors on our short-term results is performed here by studying how the changing initial
       parameters of the test orbits influence the variation of the osculating orbital elements over time. Two additional sets of 100 
       shorter control simulations are discussed here. As in the previous calculations, the initial orbital elements of each control orbit 
       have been computed varying them randomly, within the ranges defined by their mean values and standard deviations. For example, a new 
       value of the orbital eccentricity has been found using the expression $e_{\rm t} = \langle{e}\rangle + n \ \sigma_{e}\,r_{\rm i}$, 
       where $e_{\rm t}$ is the eccentricity of the test orbit, $\langle{e}\rangle$ is the mean value of the eccentricity from the available 
       orbit (Table \ref{elements}), $n$ is a suitable integer (for instance, 3, 6 or 9), $\sigma_{e}$ is the standard deviation of $e$ 
       (Table \ref{elements}), and $r_{\rm i}$ is a (pseudo) random number with normal distribution in the range $-$1 to 1.
%
%
    \begin{figure*}
      \centering
       \includegraphics[width=0.325\linewidth]{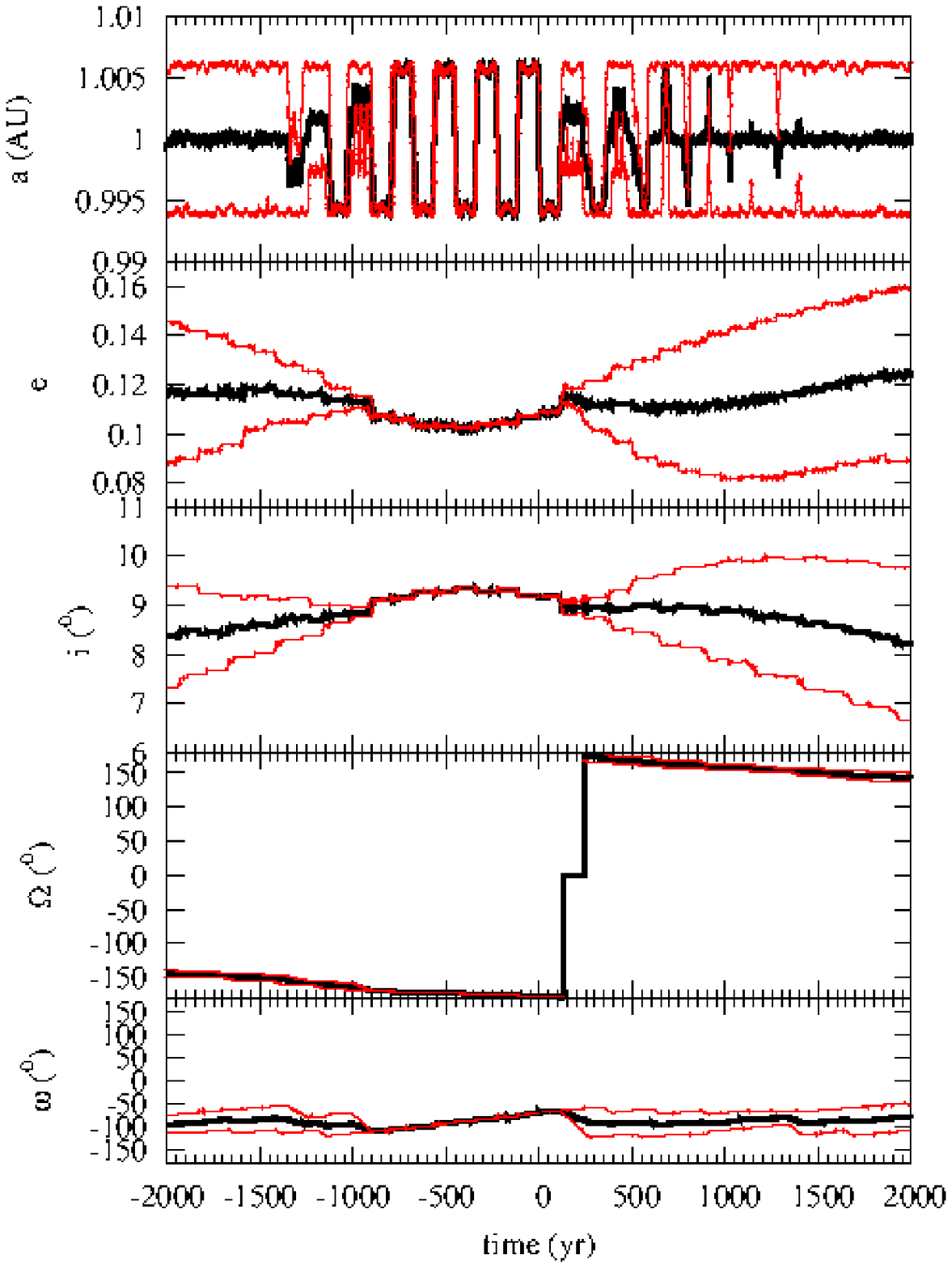}
       \includegraphics[width=0.325\linewidth]{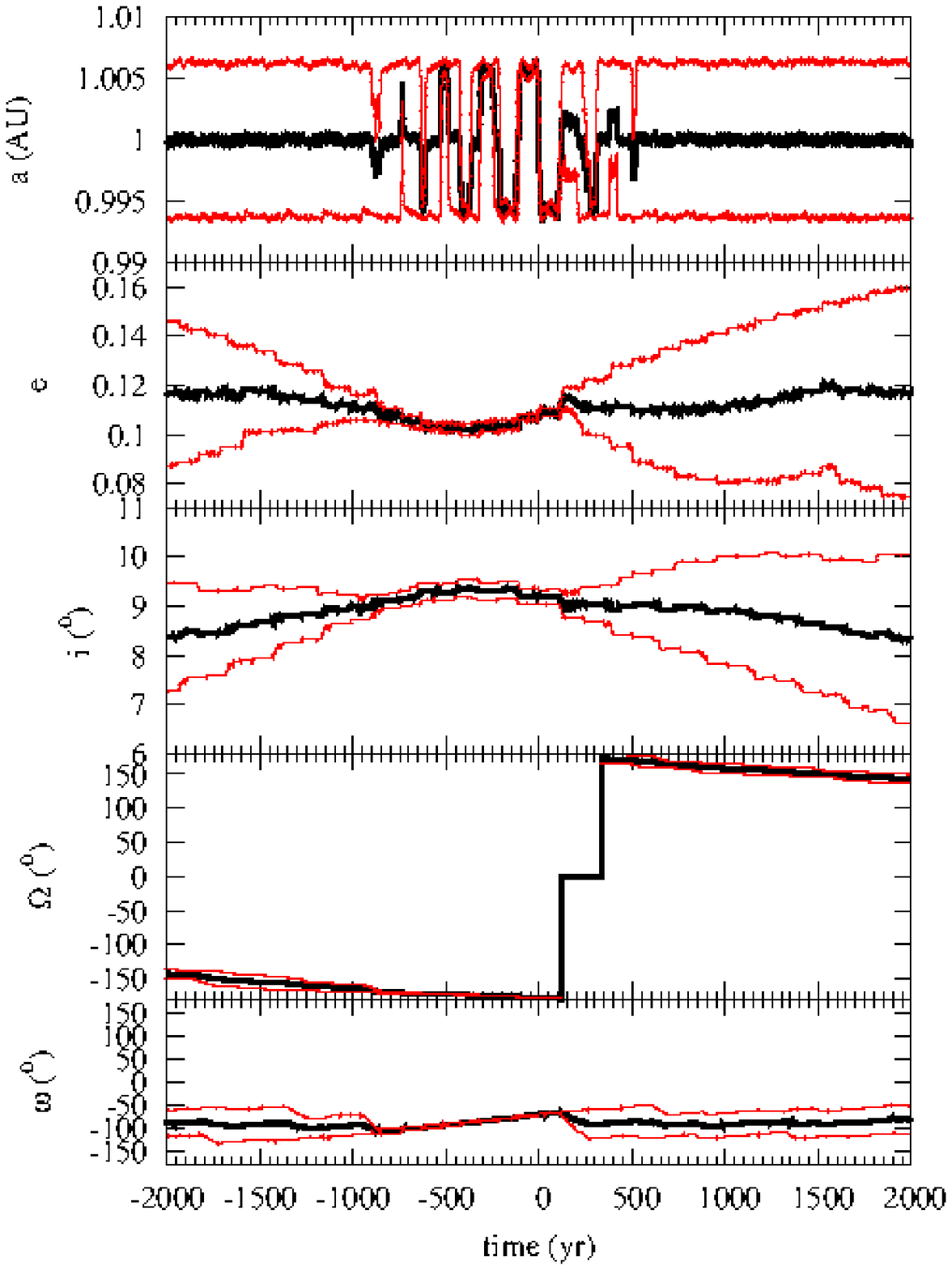}
       \includegraphics[width=0.325\linewidth]{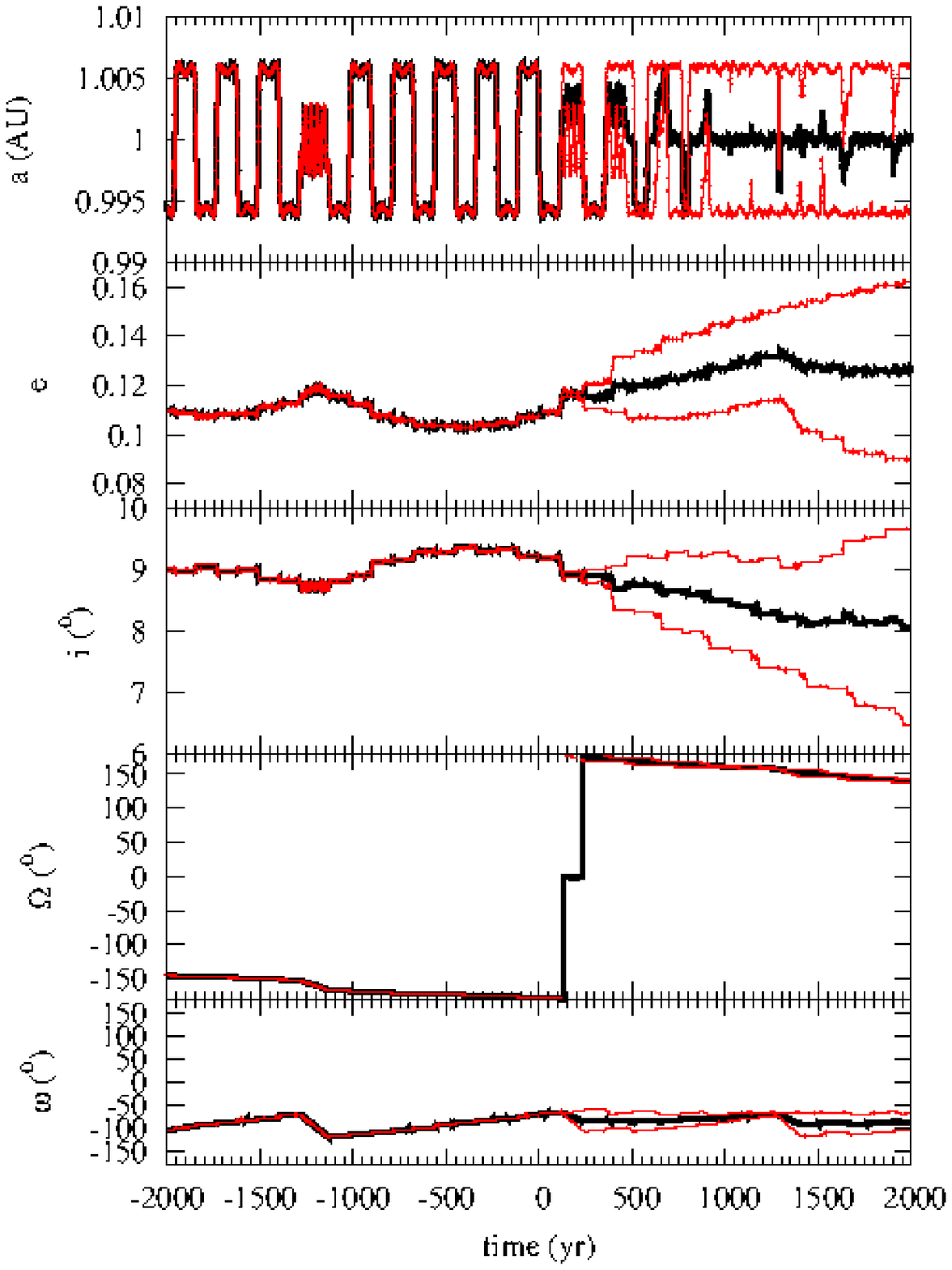}
       \caption{Time evolution of the orbital elements $a$, $e$, $i$, $\Omega$ and $\omega$ of 2015~SO$_{2}$. The black thick curve shows 
                the average evolution of 100 control orbits, the red thin curves show the ranges in the values of the parameters at the 
                given time. Results for a 1$\sigma$ spread in the initial values of the orbital elements (left-hand panels), a 6$\sigma$ 
                spread (central panels), and using MCCM (see the text, right-hand panels).
               }
       \label{errors}
    \end{figure*}
%
%

       Figure \ref{errors} shows the short-term evolution of the orbital elements $a$, $e$, $i$, $\Omega$ and $\omega$ of the object studied 
       here. The thick black curves show the average results of the evolution of 100 control orbits computed as described above. The thin 
       red curves show the ranges (minimum and maximum) in the values of the parameters at a given time. Figure \ref{errors}, left-hand and 
       central panels, shows the results for $1\sigma$ and $6\sigma$ spreads, respectively, in the initial values of the orbital elements 
       and confirms once more that even if our study is based on a relatively short-arc orbit, its results are robust and reliable. Although 
       our conclusions rest on a 9-day observational arc for 2015~SO$_{2}$, we can reliably state that this object follows a horseshoe 
       trajectory with respect to the Earth. In general, Fig. \ref{errors} shows that the value of the Lyapunov time of 2015~SO$_{2}$ is 
       very asymmetrical (about 200 yr versus several hundred years); the orbital evolution is significantly more chaotic into the future 
       than it was in the past.

    \subsection{Impact of errors: MCCM treatment}
       Sitarski (1998, 1999, 2006) has pointed out that the procedure used above is equivalent to considering a number of different virtual 
       minor planets moving in similar orbits, but not a sample of test orbits incarnated from a set of observations obtained for a single 
       minor planet. The correct statistical alternative is to consider how the elements affect each other, applying the Monte Carlo using 
       the Covariance Matrix (MCCM) approach (Bordovitsyna et al. 2001; Avdyushev \& Banschikova 2007), or to follow the procedure described 
       in Sitarski (1998, 1999, 2006).

       As a consistency test, we have used an implementation of the MCCM approach to recompute the orbital evolution of this object 
       generating control orbits with initial parameters from the nominal orbit adding random noise on each initial orbital element making 
       use of the covariance matrix (for details, see de la Fuente Marcos \& de la Fuente Marcos 2015b). Figure \ref{errors}, right-hand 
       panels, shows the results of these simulations and, in general, the difference is not very significant. The Lyapunov time asymmetry
       is also confirmed.

 \section{Discussion}
    If we compare the orbital evolution of 2015~SO$_{2}$ as depicted in Figs. \ref{first}, \ref{second} and \ref{third} with that of other
    horseshoe librators of the Earth, for instance 2001~GO$_{2}$, 2002~AA$_{29}$, and 2003~YN$_{107}$ (Brasser et al. 2004) or 
    2013~BS$_{45}$ (de la Fuente Marcos \& de la Fuente Marcos 2013) one rapidly realises that 2015~SO$_{2}$ is far more stable. Given the
    fact that the values of the semi-major axis of these objects are very similar, the source of the enhanced stability must be in the
    values of other orbital elements. Asteroids 2001~GO$_{2}$, 2003~YN$_{107}$ and 2013~BS$_{45}$ all move in orbits with $i<5\degr$; 
    2002~AA$_{29}$ has $i=10\fdg75$, but its eccentricity is rather low, 0.013. In contrast, 2015~SO$_{2}$ has $e$ = 0.11 and $i$ = 9\fdg2.
    This combination of values of $e$ and $i$ appears to be particularly favourable regarding orbital stability. However, it places the
    object far from the Earth even at close approaches; in nearly two centuries around the current epoch, the distance of closest approach 
    is 0.037 AU. If objects like 2015~SO$_{2}$ are numerous, then they will be intrinsically more difficult to discover than those similar 
    to 2013~BS$_{45}$ that reach perigee at 0.013 AU. Asteroid 2013~BS$_{45}$ is far less stable than 2015~SO$_{2}$, compare Fig. 
    \ref{first} with fig. 2 in de la Fuente Marcos \& de la Fuente Marcos (2013).

    Almost certainly, the most stable known Earth's co-orbital is 2010~SO$_{16}$ that stays as horseshoe librator for at least 120 kyr and 
    possibly up to 1 Myr (Christou \& Asher 2011), remaining in the same co-orbital configuration during this time span. Asteroid 
    2015~SO$_{2}$ may be almost as stable as 2010~SO$_{16}$ but it often switches co-orbital configuration. In addition, the value of its 
    distance from the Earth at perigee remains under 0.5 AU for about 22 yr when it reaches the ends of its horseshoe orbit, every 113 yr. 
    During its current stint as neighbour of our planet, its final perigee at $<0.5$~AU will be on 2026 September 16. In other words, 
    objects like 2015~SO$_{2}$ are characterised by relatively long favourable accessibility windows making the implementation of a 
    hypothetical space mission easier. On the other hand, our analysis leaves the question of the possible origin of 2015~SO$_{2}$ open. It 
    is clearly a transient co-orbital to our planet, but its dynamical behaviour is quite unusual. An origin in the Earth-Moon system cannot 
    be entirely ruled out based on the available information alone; its orbital evolution was significantly more stable in the past.

    Finally, it can be argued that presenting this object here is rather premature because its orbit is still not well known (see Table 
    \ref{elements}). However, this object is only directly perturbed by the Earth--Moon system and the time interval between closest 
    approaches to the Earth--Moon system is currently 113 yr. It means that the characteristic time-scale between favourable visibility 
    windows is over a century and that its orbit (see above) is quite stable. If one misses the opportunity of studying the object during
    the next few years, waiting for over a hundred years will be required to recover it. In addition, the analysis of the influence of 
    errors on our conclusions clearly indicates that our study is justified and its conclusions robust. One of the objectives of this 
    research is to bring this interesting minor body to the attention of the astronomical community, encouraging follow-up observations. 
    Spectroscopic studies during its next perigee should be able to confirm if an origin in the Earth-Moon system is plausible.

 \section{Conclusions}
    In this research, we have used $N$-body simulations and statistical analyses to study the orbital evolution of 2015~SO$_{2}$. The main 
    conclusions of our study can be summarised as follows:
    \begin{itemize}
       \item Asteroid 2015~SO$_{2}$ currently follows a horseshoe trajectory with respect to the Earth (probability $>99.9$\%), the ninth 
             asteroid known to do so.
       \item Asteroid 2015~SO$_{2}$ is a transient co-orbital that experiences a rather peculiar orbital evolution characterised by 
             alternating horseshoe and quasi-satellite episodes.
       \item Asteroid 2015~SO$_{2}$ is subjected to a Kozai resonance; its argument of perihelion oscillates around $-90\degr$ or 270\degr.
       \item Asteroid 2015~SO$_{2}$ may have had its origin in the Earth-Moon system and it is one of the most stable Earth horseshoe 
             librators discovered to date.
    \end{itemize}

 \acknowledgments
    We thank two anonymous referees for their constructive and helpful reports, and S.~J.~Aarseth for providing the main code used in this 
    research. In preparation of this paper, we made use of the NASA Astrophysics Data System, the ASTRO-PH e-print server and the MPC data 
    server.

\end{document}